\documentclass[prl,reprint]{revtex4-1}

\usepackage{graphicx}
\usepackage{bm}
\usepackage{hyperref}
\usepackage{color}

\begin{document}

\title{Capacitive Response of Wigner Crystals at the Quantum Hall
  Plateau}

\author{Lili Zhao}
\affiliation{International Center for Quantum Materials,
Peking University, Haidian, Beijing, China, 100871}
\author{Wenlu Lin}
\affiliation{International Center for Quantum Materials,
Peking University, Haidian, Beijing, China, 100871}
\author{Y.J.\ Chung}
\affiliation{Department of Electrical Engineering,
Princeton University, Princeton, New Jersey 08544}
\author{K.W.\ Baldwin}
\affiliation{Department of Electrical Engineering,
Princeton University, Princeton, New Jersey 08544}
\author{L.N.\ Pfeiffer}
\affiliation{Department of Electrical Engineering,
Princeton University, Princeton, New Jersey 08544}
\author{Yang Liu}
\email{liuyang02@pku.edu.cn}
\affiliation{International Center for Quantum Materials,
  Peking University, Haidian, Beijing, China, 100871}

\begin{abstract}
  In this report, we study ultra-high-mobility two-dimensional (2D)
  electron/hole systems with high precision capacitance
  measurement. We find that the capacitance charge appears only at the
  fringe of the gate at high magnetic field when the 2D conductivity
  decreases significantly. At integer quantum Hall effects, the
  capacitance vanishes and forms a plateau at high temperatures
  $T\gtrsim 300$ mK, which surprisingly, disappears at $T\lesssim 100$
  mK. This anomalous behavior is likely a manifestation that the
  dilute particles/vacancies in the top-most Landau level form Wigner
  crystal, which has finite compressibility and can host polarization
  current.
\end{abstract}


\maketitle

A strong perpendicular magnetic field $B$ quantizes the
kinetic energy of electrons/holes into a set of discrete
Landau levels. The discrete level structure gives rise to the
formation of incompressible quantum Hall liquids, an insulating phase
with vanishing longitudinal conductance and
quantized Hall conductance \cite{theQHE, PQHE,
  Jain.CF.2007}. Another insulating phase appears when the Landau
level filling factor $\nu=nh/eB$ is small
\cite{Jiang.PRL.1990, Goldman.PRL.1990}, which is generally believed
to be a Wigner crystal pinned by the small but ubiquitous disorder
potential \footnote{See articles by H. A. Fertig and by M. Shayegan,
  in \emph{Perspectives in Quantum Hall Effects}, Edited by S. Das
  Sarma and A. Pinczuk (Wiley, New York, 1997).}. In state-of-the-art
high-mobility 2D systems, Wigner crystals are also seen near integer $\nu=\text{N}$ 
(N is a positive integer) when the particles/vacancies in the
topmost Landau level have sufficiently low effective filling factors
$\nu^*=|\nu-\text{N}|$. \cite{Chen.PRL.2003, Liu.PRL.2012, Hatke.Nat.Comm.2014}.

The capacitance is of great interest in quantum measurements
\cite{PhysRevLett.21.212, PhysRevB.9.4410, PhysRevB.32.2696,
  MOSSER19865, PhysRevLett.68.3088, PhysRevB.34.2995,
  PhysRevLett.78.4613, PhysRevB.50.1760, nature23893, 2019,
  PhysRevLett.68.674, PhysRevLett.122.116601, PhysRevB.47.4056,
  PhysRevLett.121.167601, PhysRevLett.123.046601}.The chemical
potential $\mu$ of a Fermion system depends on its particle density,
leading to the quantum capacitance that is proportional to the density
of states at the Fermi energy. High sensitivity capacitance
measurements can reveal fine structures of the systems' energy levels,
such as the formation of delicate quantum phases \cite{MOSSER19865,
  PhysRevLett.68.3088, PhysRevB.34.2995, PhysRevLett.78.4613,
  PhysRevB.50.1760, nature23893}, the non-parabolic dispersion
\cite{2019}, the interaction induced negative compressibility
\cite{PhysRevLett.68.674, PhysRevLett.122.116601}, the minibands and
Fermi contour transitions in multi-band 2D materials
\cite{PhysRevB.47.4056, PhysRevLett.121.167601,
  PhysRevLett.123.046601}, etc. Unfortunately, quantitative studies
using high precision capacitive measurement at mK-temperature are
limited.

\begin{figure}[!hbp]
\includegraphics[width=0.45\textwidth]{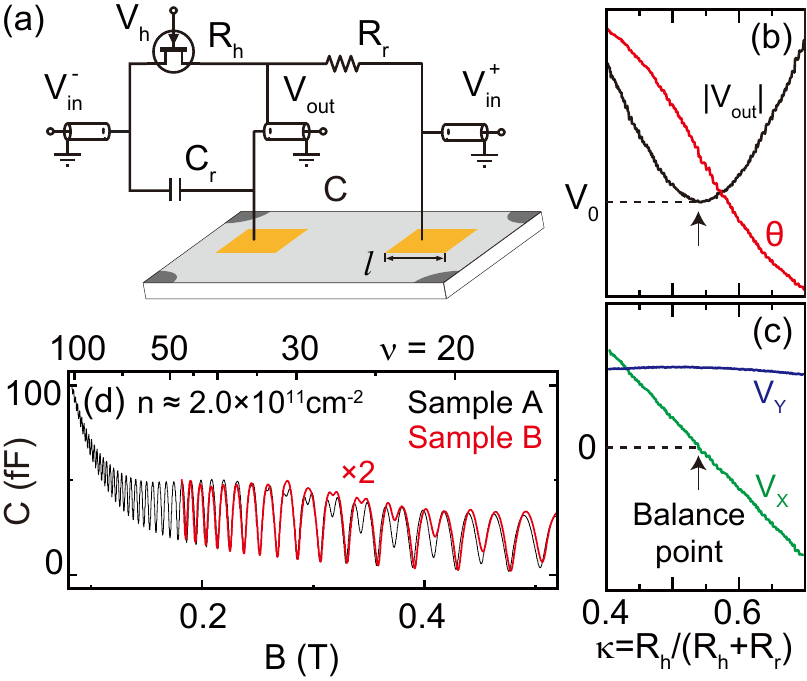}
\caption{(a) Circuit diagram of our capacitive measurement setup. (b)
  The measured amplitude and phase of output
  signal. $|V_{\text{out}}|$ reaches its minimum $V_0$ and $\theta=-\pi/2$ when the bridge
  is at balance point, indicated by the black arrows. (c)
  $V_{\text{out}}$ can be separated into the in-phase and out-of-phase
  components, $V_{\text{X}}=|V_{\text{out}}|\cdot \text{cos}\theta$
  and $V_{\text{Y}} =|V_{\text{out}}|\cdot \text{sin}\theta$. (d) The
  $C$ vs. $B$ taken from samples A (black) and B (red). The red trace
  is amplified by a factor of 2.}
\end{figure}

Here we report our high-precision capacitance studies on
ultra-high-mobility 2D electron/hole systems at mK-temperature. We
find that the device capacitance $C$
has a strong positive dependence on the 2D longitudinal sheet
conductance $\sigma$. Our observation suggests that the capacitance
charge appears only at the fringe of the gate. $C$ at
integer Landau level filling factor $\nu$ approaches zero, agrees with
the expectation that the zero $\sigma$ prohibits charge being
transported. At the vicinity of integer $\nu$, the $C$
plateau is seen as wide as the $\sigma$ plateau at $T\simeq 300$
mK. Surprisingly, it disappears at lower temperatures while the
$\sigma$ plateau becomes even wider. This anomalous behavior is likely
induced by the Wigner crystal formed by the dilute
particles/vacancies in the topmost Landau level.

The samples used in this study are made from GaAs wafers grown by
molecular beam epitaxy along the (001) direction.  These wafers
consist of GaAs quantum well bounded on either side symmetrically by
undoped AlGaAs spacer layers and $\delta$-doping layers. Samples A and
B are Si-doped electron systems with 30-nm-wide quantum well and
as-grown density $n\simeq$ 2.0 $\times 10^{11}$ cm$^{-2}$. Sample C is
C-doped hole system with 17.5-nm-wide quantum well and as-grown
density $p\simeq$ 1.6 $\times 10^{11}$ cm$^{-2}$. Each sample has
alloyed InSn or InZn contacts at the four corners of a $2\times 2$
mm$^2$ piece cleaved from the wafer. For each sample, we evaporate
several separate Au/Ti front gates with different geometries, and
measure the gate-to-gate capacitance. Samples A \& B consist 500
$\mu\text{m}$-separated square gates with $l=200$ and $100$
$\mu\text{m}$, respectively (Fig. 1(b)). Sample C consists concentric
gates (Fig. 3(b) inset). The inner gate radius is 60 $\mu\text{m}$,
and the gap between the two gates is 20 $\mu\text{m}$. We have
compared samples with different gate geometries, and find that the
observed features have no substantial dependence on the gate geometry
(data not shown). The measurements are carried out in a dilution
refrigerator with base temperature $T\simeq$ 30 mK.

Figure 1 depicts the principle of our measurement \footnote{Lili Zhao,
  \textit{et al.}, to be published}. The passive bridge installed at
the sample stage consists one resistance arm and one capacitance
arm. The resistance arm includes a reference resistance $R_{\text{r}}$
and one voltage-controlled-variable-resistance $R_{\text{h}}$,
implemented by the source-to-drain resistance of a
high-electron-mobility-transistor. We tune $R_{\text{h}}$ via the
transistor's gate voltage $V_{\text{h}}$, and measure $R_{\text{h}}$
and $R_{\text{r}}$ in-situ with low-frequency lock-in technique. We
excite the bridge with a radio-frequency voltage (typically
$\simeq 130$ MHz, $\simeq 0.5$ mV$_{\text{PP}}$). We amplify the
bridge output $V_{\text{out}}$ and measure its amplitude and phase
with a custom-built radio-frequency lock-in amplifier. The measured
$|V_{\text{out}}|$ reaches minimum value $V_0$ when the bridge is
balanced, e.g. $R_{\text{h}}/R_{\text{r}}=C/C_{\text{r}}$, see
Fig. 1(b). By properly choosing the reference phase, we can separate
$V_{\text{out}}$ into the in-phase and out-of-phase components
$V_{\text{X}}$ and $V_{\text{Y}}$. $V_{\text{X}}=0$ when the bridge is
balanced and has a linear dependence on
$\kappa=R_{\text{h}}/(R_{\text{h}}+R_{\text{r}})$ with slope
$S=\partial V_{\text{X}}/\partial \kappa$, see Fig. 1(c).
$ V_{\text{Y}}\simeq V_0$ is nearly independent on $\kappa$. At the
vicinity of the balance point, we can also deduce the $C$ from the
approximation
$V_{\text{X}}=-S\cdot(\frac{C}{C+C_{\text{r}}} - \kappa)$, which
agrees with the value measured by balancing the bridge, see Fig 2(b).

\begin{figure}[!hbp]
\includegraphics[width=0.45\textwidth]{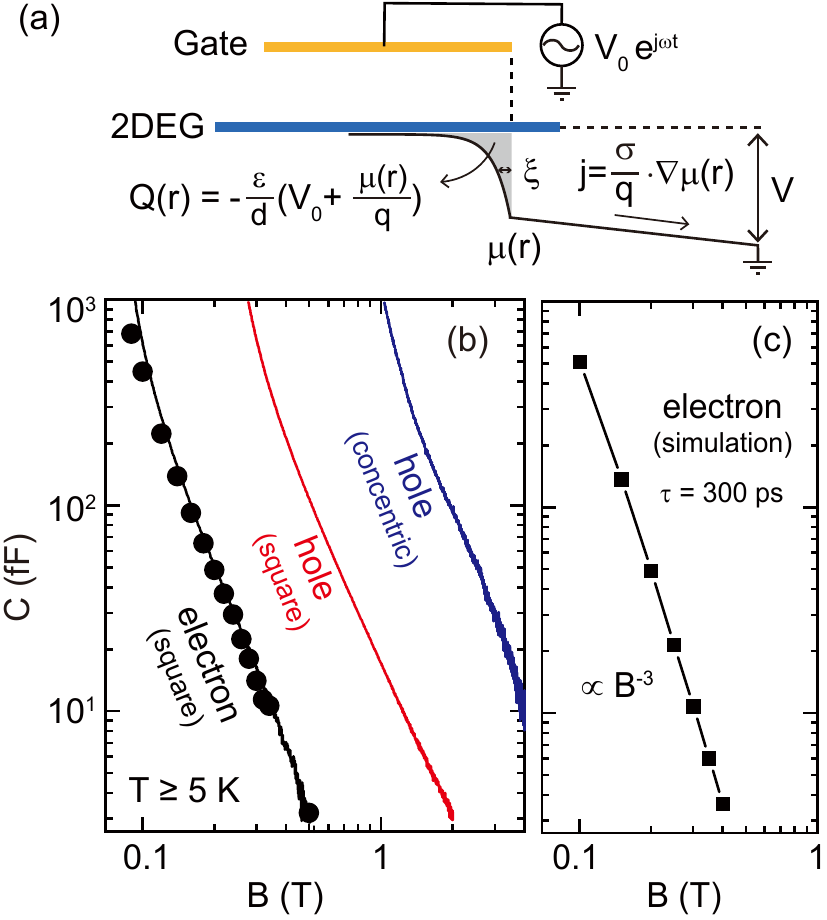}
\caption{(color online) (a) The model describing the capacitance
  response of our device. The AC voltage applied to the gate varies
  the chemical potential $\mu(\mathbf{r})$ of the underlying 2D
  system, inducing a capacitance charge density $Q(\mathbf{r})$ and
  corresponding current density $\mathbf{j}(\mathbf{r})$. (b)
  $C$ measured from three different samples at high
  temperatures $T\ge$ 5K. The symbols represents $C$
  deduced from the balance condition and the lines are results
  converted from $V_{\text{X}}$. (c) Numerical simulation predicts the
  $C\propto B^{-3}$ dependence. We use $\tau$=300 ps to
  match with the Fig. 2(b) electron data.}
\end{figure}

\begin{figure*}[!htbp]
\includegraphics[width=0.9\textwidth]{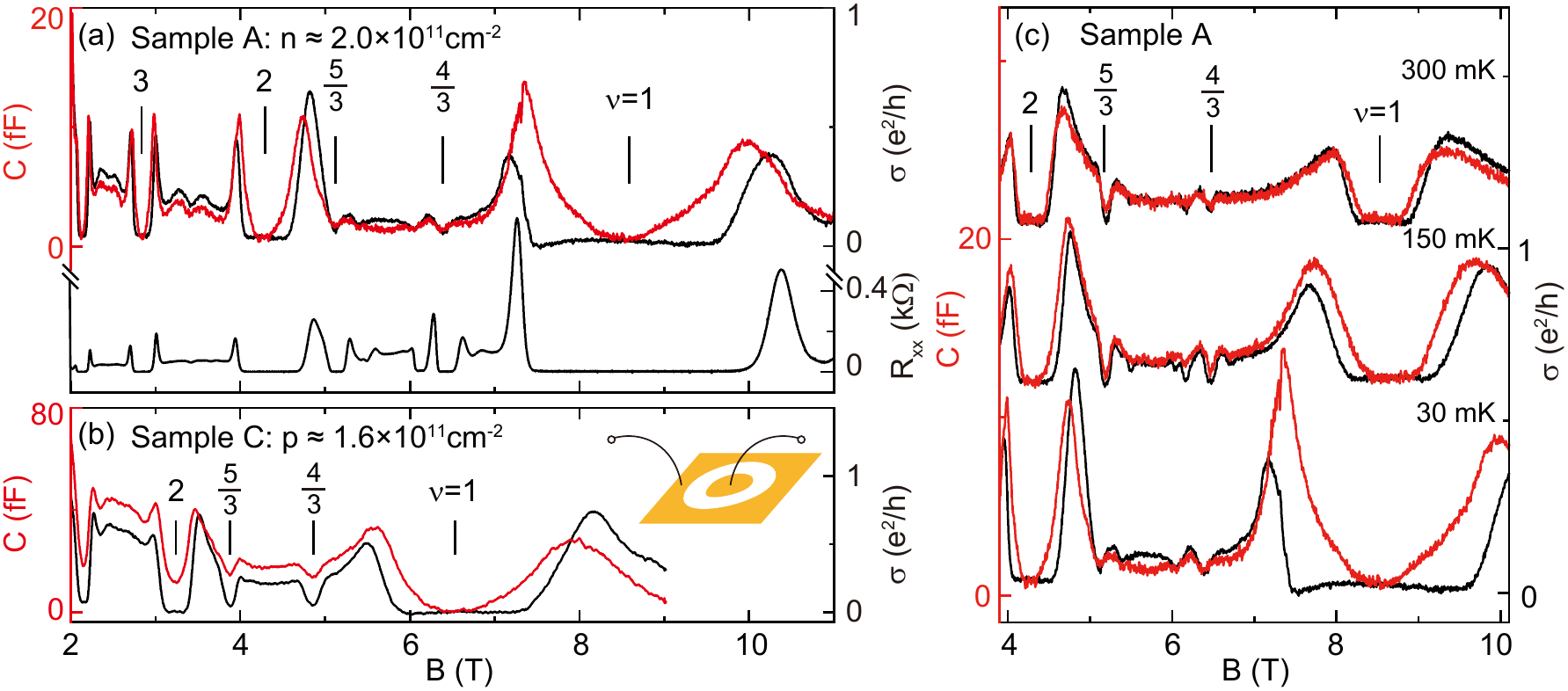}
\caption{(color online) (a) $C$, $\sigma$ and the 
  longitudinal resistance $R_{xx}$ taken from sample A. (b)
  $C$, $\sigma$ taken from sample C. (c) $C$
  and $\sigma$ taken from sample A at different temperatures. Traces
  are offset vertically.}
\end{figure*}

The $C$ data in Fig. 1(d) appreciates the merit of our
high precision measurement. As $B$ increases, $C$
decreases dramatically from its $B=0$ value (which is close to the
estimated geometric capacitance of a few pF) by orders of magnitude
\footnote{In all samples, our measured capacitance approaches a
  constant value $\simeq$60 fF when the particles forms incompressible
  integer quantum Hall liquid. This is likely the parasitic
  capacitance $C_{\text{P}}$ induced by the bonding wires, gates,
  etc. We have subtracted this value in all figures.}. Similar
phenomenon has been reported in previous experiments where the
screening capability of high quality 2D reduces significantly at high
field \cite{PhysRevLett.122.116601}. The shrinking of $C$
is less violent in samples which has shorter scattering time or
smaller size, or when we use lower measurement frequencies, also consistent
with other studies \cite{PhysRevB.32.2696, MOSSER19865,
  PhysRevB.34.2995}. In Fig. 1(d), we compare data taken from samples
A \& B, whose gate dimensions differ by a factor of 2 and center-to-center
distances are kept the same. In both samples, the capacitance oscillation
starts at $B\lesssim 0.01$ T when $\nu\gtrsim 150$, evidencing that
our measurement is as gentle as DC transport. Interestingly, the
traces taken from two samples are nearly a replica of each other but
scales by a factor of 2.

We can understand these features with the model shown in
Fig. 2(a). The 2D system is grounded through contacts remote from the
gates. A time-varying voltage $V_0\cdot e^{i\omega t}$
applied on the gate induces an oscillating capacitance charge density
$Q(\mathbf{r}) \cdot e^{i\omega t}$ in the 2D system satisfying
$Q(\mathbf{r})= -\frac{\varepsilon}{d} (V_0+\mu(\mathbf{r})/q)$, where
$\varepsilon$ is the dielectric constant, $d$ is the gate-to-2D distance,
$q=\pm e$ is the particle's charge and $\mu(\mathbf{r})$ is the local
chemical potential of the 2D system. For simplicity, we neglect the
time-dependence term $e^{i\omega t}$ and replace $\partial/\partial t$
with $i\omega$ in the following text. The spatial variation of
$\mu(\mathbf{r})$ generates a current distribution
$\mathbf{j}(\mathbf{r}) = \frac{\sigma} {q} \nabla \mu(\mathbf{r})$ in
the 2D system and by the charge conservation law,
$\nabla\cdot \mathbf{j}(\mathbf{r}) = i\omega\cdot Q(\mathbf{r})$.

At high field when $\omega_{\text{C}}\tau\gg 1$, $\omega_{\text{C}}$
is the particles' cyclotron frequency and $\tau$ is their scattering
time, the $\sigma$ of an ultra-high mobility 2D system vanishes as
$\sigma\simeq \sigma_{(B=0)}/(1+(\omega_{\text{C}}\tau)^2)$.
$\nabla \mu(\mathbf{r})$ is almost zero except at the proximity of the
gate boundary, and $\mu(\mathbf{r})\approx qV_0$ at the center of the
gate. Near this edge, $Q(\mathbf{r})\propto\pm\exp (-|d|/\xi)$ where
$d$ is the distance from the boundary and
$\xi=\sqrt{\sigma d/(\omega \varepsilon)}$. Furthermore, the current
induced potential change outside the gated region as well as the
parasitic capacitance $C_{\text{P}}$ become non-negligible when
$\sigma$ is small. Combining all above effects, the
$C=\int Q(\mathbf{r}) d \mathbf{r} /V_0$ is proportional
to the length of the gate perimeter and reduces as $B^{-3}$, see the
numerical results in Fig. 2(c). We show $C$ measured in
2D electron and hole samples with different gate geometries in
Fig. 2(b). The measured $C$ is in nearly perfect
agreement with the $B^{-3}$ prediction. This model also predicts that
$C$ decreases slower if $\tau$ is smaller or the
effective mass is larger, also consistent with our observation in Fig.
2(b) where $C$ in 2D hole systems is usually higher than
in 2D electron systems at high field.

Based on the Fig. 2 model, the out-of-phase signal $V_{\text{Y}}$ is a
good measure of $\sigma$, where
$V_{\text{Y}}=V_0-S\cdot(\frac{\sigma}{\sigma+\omega C_{\text{r}}}
-\kappa)$ if $\omega_{\text{C}}\tau\gg 1$. Fig. 3(a) shows the deduced
$\sigma$ and the $R_{xx}$ of sample A obtained from the in-situ
quasi-DC transport measurement using the four corner contacts. Both
$\sigma$ \& $R_{xx}$ traces have wide, flat plateaus when the 2D
system form incompressible integer quantum Hall insulator at $\nu=1$,
2, 3, etc. We also notice that $C$ and $\sigma$ are
remarkably similar with each other. This may not be surprising since
the Fig. 2 model predicts that $C$ strongly entangles
with $\sigma$. Mysteriously, no plateau is seen in the
$C$ trace at integer fillings while one would expect
vanishing $C$ when $\sigma=0$. The missing plateau in
$C$ is substantiated by the Fig. 3(b) data, taken from
sample C which is a 2D hole system with concentric gate geometry.

\begin{figure}[!hbp]
\includegraphics[width=0.45\textwidth]{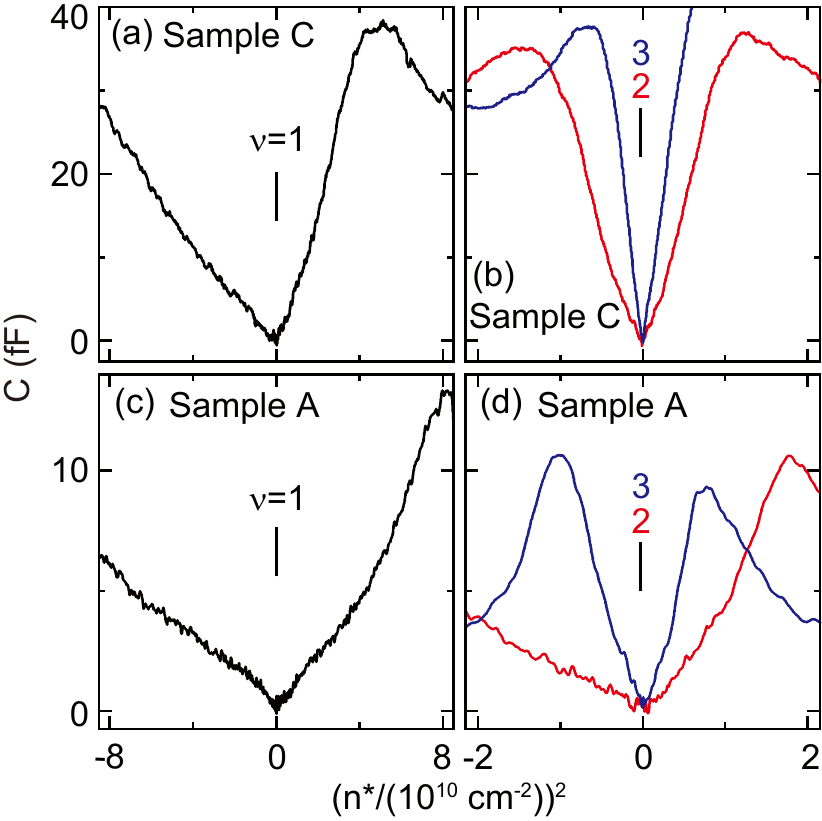}
\caption{(color online) (a \& b) $C$ taken from sample C
  near $\nu=1$, 2 and 3 as a function of $(n^*)^2$.
  The negative $(n^*)^2$ represents data from $\nu^*<1$
  while the positive $(n^*)^2$ is data from $\nu^*>1$. (c \& d)
  $C$ vs. $(n^*)^2$ near $\nu=1$, 2 and 3, taken from
  sample A.}
\end{figure}

The Fig. 3(c) data taken from sample A at 30, 150 and 300 mK is even
more intriguing. At the highest temperature $T\simeq 300$ mK, the
$C$ and $\sigma$ traces are almost exact replica of each
other. Both of them have well-developed plateau at $\nu=1$ and 2, a
manifestation that the 2D system form quantum Hall insulator and the
extra particles/holes at the vicinity of $\nu=1$ and 2 are localized
and cannot response to external electric field. When the 2D system
becomes colder, the $C$ remains zero at exact integer
fillings $\nu=1$ and 2, but its plateau gradually becomes narrower and
eventually disappears at $\simeq 30$ mK while the $\sigma$ plateau
broadens.

The fact that $C$ becomes finite at low temperatures while $\sigma=0$
suggests that charge can still be effectively transported in-plane
while the 2D system is not conducting. Plateaus appear near $\nu=1$
and 2 in the high-temperature trace when the system exhibits
incompressible quantum Hall insulators, because that the extra
particles/vacancies are localized and cannot response to the AC
voltage. Surprisingly, $C$ becomes finite at low temperature when the
localization is expected to be stronger. One plausible explanation is
the existence of a long-range correlated compressible phase which is
stable only at low temperatures. It has been suggested by previous
studies that the dilute particles/vacancies in the topmost Landau
level may form a Wigner crystal at the vicinity of integer filling
factors \cite{Chen.PRL.2003}. This solid phase cannot host conducting
current, but its deformation in time-varying external electric field
generates polarization current. The finite $C$ is an outcome of the
finite compressibility of this Wigner crystal. In such a senario, $C$
approaching zero at $T \gtrsim 200$ mK signals the melting of the
Wigner crystal.

We investigate the capacitive response of the Wigner crystal
quantitatively in Fig. 4. Figs. 4(a) \& (b) show data taken from the
sample C near $\nu=1$, 2 and 3. The capacitance $C$ has a linear
dependence on $(n^*)^2$, where $n^*=\frac{\nu^*}{\nu}n$ is the density
of particles/vacancies that form the Wigner crystal. Such linear $C$
vs. $(n^*)^2$ dependence is also seen in the electron sample near
$\nu=2$ and 3 (Fig. 4(d)). However, data taken from the electron
sample near $\nu=1$ has a skewed '$\nu $'-shape (Fig. 4(c)). The could
be possibly related to electrons' small spin splitting. We also notice
that the $C$ vs. $(n^*)^2$ in Fig. 4(a) is quite asymmetric for
positive and negative $n^*$, while the traces in Fig. 4(b) are
more-or-less symmetric. The slope $|\partial C /\partial ((n^*)^2)|$
at positive $n^*>0$ is about 3-times larger than the negative $n^*>0$
side. This might because that the $n^*>0$ ($n^*<0$) Wigner crystals
are formed by particles (vacancies) in different Landau levels, and
have different compressibility. The intriguing experimental
observations of Wigner crystals in Fig. 4 call for future theoretical
and experimental investigations.

In conclusion, with the help of our high-precision capacitance
measurement setup, we carefully study the capacitance response of
ultra-high mobility 2D systems at mK temperature. Our result shows
that the device capacitance strongly entangles with the 2D
conductivity $\sigma$ and vanishes if $\sigma=0$. Surprisingly, at the
$\sigma=0$ plateau of integer quantum Hall effects, $C$ is only zero
at high temperatures but becomes finite at base temperature. This
anomalous behavior is consistent with the formation of compressible
Wigner crystal which can response to AC voltage through polarization
current. Our experimental result suggests that this response depends
on the Wigner crystal density $n^*$ as $C\propto (n^*)^2$.

\begin{acknowledgments}
  We acknowledge support by the NSFC (Grant No. 92065104 and 12074010)
  for sample fabrication and measurement, and the NSF (Grants
  DMR-1305691, and MRSEC DMR-1420541), the Gordon and Betty Moore
  Foundation (Grant GBMF4420), and Keck Foundation for the material
  growth. We thank M. Shayegan, Xin Wan, Bo Yang, Wei Zhu
  for valuable discussion.
\end{acknowledgments}

\bibliography{bib_full}

\end{document}